\documentclass[final,5p,times,twocolumn]{elsarticle}

\usepackage{natbib}
\usepackage{array}
\usepackage{booktabs}
\usepackage{tabu}
\usepackage{dcolumn}
\usepackage{amsmath}
\usepackage{amsfonts}
\usepackage{amssymb}
\usepackage{graphicx}
\usepackage{subfigure}
\usepackage{epsfig}

\usepackage{dcolumn}
\usepackage{bm}

\def\be{\begin{equation}}
\def\ee{\end{equation}}
\def\bea{\begin{eqnarray}}
\def\eea{\end{eqnarray}}

\journal{Physics Letters B}

\begin{document}

\begin{frontmatter}



\title{An exact solution for a force-free field electrodynamics accretion disk surrounding a perturbed charged black hole}


\author[NWU,IPM,IASBS]{Haidar Sheikhahmadi\corref{cor1}}\ead{h.sh.ahmadi@gmail.com;h.sheikhahmadi@ipm.ir}

\cortext[cor1]{Corresponding author}
\address[NWU]{Center for Space Research, North-West University, Mafikeng, South Africa}
\address[IPM]{School of Astronomy, Institute for Research in
Fundamental Sciences (IPM),  P. O. Box 19395-5531, Tehran, Iran}
\address[IASBS]{Institute for Advanced Studies in Basic Sciences (IASBS), Gava Zang, P. O. Box 45137-66731, Zanjan,  Iran}

\begin{abstract}
Recently the first real image of the supermassive black hole at the heart of the galaxy M87 successfully has been captured. To explain the behavior of both the observed powerful jets and the electromagnetic accretion disk the Blandford–Znajek mechanism \cite{EHT1}, by the Event Horizon Telescope collaboration, has been considered. To justify these phenomena we are seeking an exact solution for a perturbed Rissner-Nordstr\"{o}m black hole surrounded by a force-free field electrodynamics accretion disk. Although a charged black hole is a toy model, by tending the charge to zero one can obtain the results for a perturbed, rotating, Schwarzschild one immediately. Finding an exact solution in regards to the force-free field electrodynamics as a source for perturbations in different classes of the black holes is may one of the open problems yet. Here to find the solution,  different Maxwell's scalars besides the energy-momentum tensor in both tangent and bend backgrounds are calculated. In doing so, the well-known Newman-Penrose formalism is considered \cite{NP}.
\end{abstract}


\end{frontmatter}


\section{Introduction}

\emph{Seeking an  exact solution for a perturbed Rissner-Nordstr\"{o}m metric, is the main goal of doing this work. To do so, it is assumed that the perturbations are originated by an accretion disk containing a  force-free field electrodynamics}\\
Although the physics of Force-Free Field, FFF, electrodynamics was introduced before \cite{ChandraBookHydro}, it was in $1969$ and then in $1977$ that for the first time the FFF concept  utilized to the pulsars \cite{J1} and active galactic nuclei \cite{J2}, respectively. In \cite{EHT1,J2}, it has shown that, under some specific conditions, a Black Hole, BH, can lost its rotational energy, even more angular momentum, into the magnetosphere through some electromagnetic mechanisms. They showed that as long as  the BH is surrounding by an accretion disk containing a FFF, then one can observe  the outflowing currents for instance as powerful jets. In other words, such flow can carry energy and angular momentum away from the BH to the disk. One can see a vast amount of seminal papers have devoted to the understanding  the physics of the solutions for rotating BHs enriched by the FFF electrodynamics \cite{Ref16J,Ref16Ja,Ref16Jb,Ref16Jc,Ref16Jd,Ref16Je}. Amongst such remarkable studies, one would refer to the explanation of the jets emanating from the BHs \cite{Ref16J,Ref16Ja,Ref16Jc}. Recently Jacobson and his colleagues \cite{Jac} by virtue of the FFF electrodynamics  have proposed some general solutions for rotating BHs, especially the well-known Kerr metric. They mentioned precisely finding some  solutions for such setups although do exist, all of them are numeric not exact analytical ones. Even more, to find the exact form of the solutions have presented in \cite{Jac} one has to introduce, i.e. as an ansatz for instance, at least  the form of  the  separable functions in one side or the total current on the other side to find the general solution.\\
 To cope with aforementioned ambiguities following the null tetrad formalism and by means of the perturbation method we will seek an exact analytical solution for a  rotating Rissner-Nordstr\"{o}m, R-N BH \cite{RN,RNa} with an electrodynamics accretion disk \cite{ChandraPapers-RN paperg,ChandraPapers-RN paper,ChandraPapers-RN papera,ChandraPapers-RN paperb,ChandraPapers-RN paperc,ChandraPapers-RN paperd,ChandraPapers-RN papere,ChandraPapers-RN paperf}. It should be noted, in such perturbed background one can separate the perturbations into axial and polar categories.

\section{Perturbed metric in the Newman-Penrose approach}
  For this investigation we consider a covariant form of the metric which allows for the non-stationary as follows:
 \begin{eqnarray}\label{metric}
 d{S^2} = {e^{2\nu }}{(dt)^2} - {e^{2\psi }}{(d\varphi  - \omega dt - {\tilde{q}_2}d{r} - {\tilde{q}_3}d{\theta})^2}\\ \nonumber
 ~~~ - {e^{2{\mu _2}}}{(d{r})^2} - {e^{2{\mu _3}}}{(d{\theta})^2}~,
 \end{eqnarray}
 where the introduced seven parameters $\nu,~\psi,~\omega,~\mu_2,~\mu_3,~\tilde{q}_2$, and $\tilde{q}_3$ are some functions of $r$, $\theta$, and time as well. It should be noted the Einstein's equation is covariant and provides six independent equations for the metric elements and consequently these seven parameters will be occurred only in six independent combinations. The basis vectors $(l,~n,~m,~\bar m)$ to construct the null tetrad frame based on R-N metric in a N-P formalism could be expressed as follows:
 \begin{eqnarray}\label{Contra}
\begin{array}{l}
{l^p} = ({l^t},{l^r},{l^\theta },{l^\varphi }) = (\frac{{{r^2}}}{\Delta },1,0,0),\\
{n^p} = ({n^t},{n^r},{n^\theta },{n^\varphi }) = (1, - \frac{{{r^{ - 2}}}}{{2{\Delta ^{ - 1}}}},0,0),\\
{m^p} = ({m^t},{m^r},{m^\theta },{m^\varphi }) = (0,0, - \frac{{{r^{ - 1}}}}{{\sqrt 2 }},\frac{{i{r^{ - 1}}}}{{\sqrt 2 }}\csc \theta )~,
\end{array}
\end{eqnarray}
where $\Delta=r^2-2Mr+Q^2=r^2e^{2\nu}=r^2e^{-2\mu_2}$, $l \cdot n = 0,\,\,\,m \cdot \bar m =  - 1$, and $\bar m$ is the conjugate of $m$. We denoted by $M$ total mass of the BH and by $Q$ the charge of the in question BH. Based on the definition of the metric, (\ref{metric}), the contravariant orthonormal basis of the model can be obtained as
 \begin{eqnarray}\label{ContrBasis}
\begin{array}{l}
e_0^a = ({e^{ - \nu }},\,\omega {e^{ - \nu }},\,0,\,\,\,0)~,\\
e_1^a = (0,\,\,{e^{ - \psi }},\,\,\,\,0,\,\,\,\,\,\,\,\,0)~,\\
e_2^a = (0,\,{\tilde{q}_2}{e^{ - {\mu _2}}},\,{e^{ - {\mu _2}}},\,0)~,\\
e_3^a = (0,\,{\tilde{q}_3}{e^{ - {\mu _3}}},\,0,\,{e^{ - {\mu _3}}}).
\end{array}
\end{eqnarray}
 Before going any further, the perturbations in the metric can be discomposed   into the axial and the polar types. Despite the polar ones, in the axial portion of the perturbations under changing the sign of variable $\varphi$ the sign of the perturbed terms in the metric undergoes a change in sign  and consequently one can expect a rotation or asymmetry in the BH. In other words, obviously, in the axial part the parameters $\omega,~\tilde{q}_2$, and $\tilde{q}_3$ will be affected after considering a reversal in sign for $\varphi$. These different treatments lead to this fact that the axial and polar perturbations can be appeared independently and so they should be considered as decouple subsystems. Therefore, by virtue of the perturbation approach for tetrad bases in the N-P formalism we can obtain the axial, say imaginary or odd, and the polar , say real or even, portions of the Weyl scalars namely $\Psi_0$ and $\Psi_2$ as follows: one should note that hereafter, for simplicity,  the portion of dragging by eliminating $\omega$ is neglected, i.e. $\omega=0$,
 \begin{eqnarray}\label{Weyl0Axial}
  - 2i\sigma {\mathop{\rm Im}\nolimits} {\Psi _0} = \frac{{{r^3\frac{{C_{l + 2}^{ - 3/2}}}{{{{\sin }^2}\theta }}}}}{{{{({q_1}[{q_1} - {q_2}])^2}\Delta ^2}}}({Y_{ + 2}}\cos \bar \psi  + {Y_{ + 1}}\sin \bar \psi )~,
 \end{eqnarray}
and
 \begin{eqnarray}\label{Weyl4Axial}
 - 2i\sigma {\mathop{\rm Im}\nolimits} {r^4}{\Psi _4} = \frac{{{r^3}\frac{{C_{l + 2}^{ - 3/2}}}{{{{\sin }^2}\theta }}}}{{4{{({q_1}[{q_1} - {q_2}])}^2}{\Delta ^2}}}({Y_{ - 2}}\cos \bar \psi  + {Y_{ - 1}}\sin \bar \psi )~,
  \end{eqnarray}
  where \[{Y} _{\pm j} {\text{ = }}V_j^ \pm Z_j^ \pm  + (W_j^ \pm  + 2 i \sigma ){\Lambda_ + }Z_j^ \pm,\]
  and for different definitions have appeared in above equation we receive
  $$W_{j}^{(-)}=W^{(-)}=\frac{2}{r^{2}}\left(r-3 M+2 \frac{Q^{2}}{r}\right),$$ and
   $$W_{j}^{(+)}=W^{(-)}-2 q_{k} \frac{\Delta}{r^{3}\left(\mu_C^{2} r+q_{j}\right)} \quad(j, k=1,2 ; j\neq k).$$
 where $\mu _C^2 = 2n = (l - 1)(l + 2).$
The real constant $\sigma$ was appeared because of introducing the time-dependency of the perturbations as $exp[i\sigma t]$.
Besides we obtained these solutions and have used them in the Regge-Wheeler and the Zerreli like equations respectively \cite{RW-ZE} and \cite{ZEa,ZEb}
\begin{eqnarray}\label{Zerrili}
\begin{array}{l}
Z_1^{( - )} = \sqrt {({q_1} - {q_2}){q_1}} (H_1^{( - )}\cos \bar \psi  + H_2^{( - )}\sin \bar \psi ),\,\\
Z_2^{( - )} = \sqrt {({q_1} - {q_2}){q_1}} (H_2^{( - )}\cos \bar \psi  - H_1^{( - )}\sin \bar \psi ),\,\\
Z_1^{( + )} = {q_1}H_1^{( + )} + i\sqrt {{q_1}{q_2}} H_2^{( + )},\,\\
Z_2^{( + )} =  - i\sqrt {{q_1}{q_2}} H_1^{( + )} + {q_1}H_2^{( + )}\,.
\end{array}
\end{eqnarray}
Where
\[\begin{array}{l}
\sin 2\bar \psi  = \frac{{2i\sqrt {{q_1}{q_2}} }}{{{q_1} - {q_2}}} = \frac{{2Q{\mu _C}}}{{\sqrt {{{(3M)}^2} + {{(2Q{\mu _C})}^2}} }},\,\\
{q_1} = 3M + \sqrt {{{(3M)}^2} + {{(2Q{\mu _C})}^2}} ,\,\\
{q_2} = 3M - \sqrt {{{(3M)}^2} + {{(2Q{\mu _C})}^2}} ,\, .
\end{array}\]
Accordingly we used the following abbreviations
\[\begin{array}{l}
V_j^{( - )} = \frac{\Delta }{{{r^5}}}((\mu _C^2 + 2)r - {q_j}(1 + \frac{{{q_k}}}{{\mu _C^2r}})),\,(j,k = 1,2,j \ne k)\\
\\
and\\
\\
V_1^{( + )} = \frac{\Delta }{{{r^5}}}(U + \frac{{({q_1} - {q_2})}}{2}W),\,\,V_2^{( + )} = \frac{\Delta }{{{r^5}}}(U - \frac{{({q_1} - {q_2})}}{2}W)~,
\end{array}\]
one notice that we have defined the following abbreviations
\[\begin{array}{l}
U = (2nr + 3M)W + (\varpi  - nr - M) - \frac{{2n\Delta }}{\varpi }~,\\
\\
W = \frac{\Delta }{{r{\varpi ^2}}}(2nr + 3M) + \frac{1}{\varpi }(nr + M)~,
\end{array}\]
where
\[\begin{array}{l}
\varpi  = nr + 3M - \frac{{2{Q^2}}}{r},\\
\\
\frac{1}{r} - {v_{,r}} = \frac{1}{{r\Delta }}({r^2} - 3Mr + 2{Q^2}).\,
\end{array}\]
We immediately understood  the real part of $\Psi_0$ is as same as the equation \ref{Weyl0Axial} without the coefficient $2i\sigma$ on the left hand side.  To investigate the ingoing and outgoing currents, say jets, we must find the solutions based on the null tetrads in regards of the FFF electrodynamics. But at first we can obtain all components of Maxwell's scalars for the perturbed R-N-BH and then we can consider the FFF conditions to obtain the in question exact solutions. Here automatically for $2\phi_1={F_{pq}}({{\mathop{\rm l}\nolimits} ^p}{n^q} + {m^p}{{\bar m}^q})$ we have $Q/2r^2$. Now for the remnant of Maxwell's scalars, axial and polar parts,  respectively one has
\begin{eqnarray}\label{phi0axial}
\begin{array}{l}
2i\sigma{\mathop{\rm Im}\nolimits} {\phi _0}=
\frac{ {3r}{\Lambda _ + }(Z_1^{( - )}\cos \bar \psi  - Z_2^{( - )}\sin \bar \psi )P_l,_{\theta}}{{\Delta } {\sqrt{(2{q_1}[{q_1} - {q_2}])}}}~,
\end{array}
\end{eqnarray}
\begin{eqnarray}\label{phi0Polar}
\begin{array}{l}
2\sqrt 2 {\rm{Re }}{\phi _0} = \mu_C \frac{r}{\Delta }{\Lambda _ + }(H_1^{( + )}\cos \bar \psi  + \frac{{2Q}}{{\mu r}}\Phi )P_l,_{\theta}~,
\end{array}
\end{eqnarray}
and
\begin{eqnarray}\label{phi2axial}
\begin{array}{l}
 2i\sigma{\rm{Im}}{\phi _2} = \frac{{{3r}}{}{\Lambda _ - }(Z_1^{( - )}\cos \bar \psi  - Z_2^{( - )}\sin \bar \psi  )P_l,_{\theta}}{{ \Delta }\sqrt {(2{q_1}[{q_1} - {q_2}])}}~,
\end{array}
\end{eqnarray}
where we used the relation between  Gegenbauer function and Legendre function, i.e. $C_{l + 2}^{ - 3/2}(\theta ) = {\sin ^2}\theta ({P_{l,\theta ,\theta }} - {P_{l,\theta }}\cot \theta )$. We notice the functions $Z_1^{( - )}$ and $Z_2^{( - )}$ besides $H_1^{( + )}$ can satisfy the Schr\"{o}dinger-like wave equations and under some specific constraints they will reduce to Regge-Wheeler and Zerreli equations respectively  \cite{RW-ZE} and \cite{ZEa,ZEb}.\\

\section{Main results}

 The FFF electrodynamics is related to the systems in which the stored energy and momentum in the field is much larger than the matter. To understand what such a field is one can visualize for instance a strong magnetic field contains a low density plasma. Utilizing the FFF condition, viz.,
\begin{equation}
F_{pq}J^{q}=0~,
\end{equation}
 as an extra condition besides the usual Maxwell's equations can be considered to find the solutions for the components of the 4-current. Following the N-P formalism \cite{NP} but for FFF electrodynamics we interestingly found that if one consider the current along one of the congruences, or null directions, e.g. $l^p$ or $n^p$, the complicity of the equations dramatically will be decreased. By these assumption obviously it will be realized that $\phi_1$ takes the imaginary values, and one of the scalars $\phi_2={F_{ab}}{{\bar m}^a}{m^b} = \frac{{{e^\upsilon }}}{{2\sqrt 2 }}[i({F_{01}} + {F_{21}}) + ({F_{03}} + {F_{23}})]$ or $\phi_0 = {F_{ab}}{l^a}{m^b} = \frac{{{e^{ - \upsilon }}}}{{\sqrt 2 }}[i({F_{01}} + {F_{21}}) + ({F_{03}} + {F_{23}})]\,$ based on the selection of direction will be equal to zero.  Therefore, for the ingoing solutions $J^p\propto n^p$ and we found $\phi_2=0$. For the outgoing solutions $J^p\propto l^p$ and $\phi_0=0$. These results lead to this fact that for the FFF electrodynamics because of purely ingoing or outgoing solutions  the scattering  from the surface of the BHs does not exist.  By virtue of the spin coefficient formalism of N-P approach, and for ingoing category for a rotating R-N-BH we can obtain the imaginary and real  parts of Maxwell's scalar $\phi_0$ as follows:
\begin{eqnarray}\label{phi0axial}
Im \phi_0=\frac{-ir}{\Delta \sin\theta}\Upsilon_{,\phi}(v, \theta, \varphi)~,
\end{eqnarray}
and
\begin{eqnarray}
Re \phi_0=\frac{-r}{\Delta}\Upsilon_{,\theta}(v, \theta, \varphi)~,\label{phi0Polar}
\end{eqnarray}
where $v=t+r_*$ and $\Delta$ is, as mentioned above, the horizon function. Also the ingoing 4-current based on FFF Maxwell equations can be expressed as follows
\begin{eqnarray}\label{current}
\sqrt{8}\Delta\pi J=(\cot \theta \partial_{,\theta}+\partial_{,\theta,\theta}+\frac{1}{\sin^2\theta}\partial_{,\varphi,\varphi})\Upsilon(v, \theta, \varphi)~.
\end{eqnarray}
Now by equating axial  and polar parts of perturbed solutions we can find the exact solution  for the Maxwell's scalar $\phi_0$ respectively as following:
\begin{eqnarray}\label{Upsilonaxial}
\Upsilon_{axial}(v,\theta ,\varphi ) &=& \int_{\varphi_1}^{\varphi_2}\frac{{3\sin \theta }}{{\sigma {{(2{q_1}[{q_1} - {q_2}])}^{1/2}}}}\\ \nonumber
&\times&{\Lambda _ + }(Z_1^{( - )}\cos \bar \psi  - Z_2^{( - )}\sin \bar \psi ){P_{l,\theta }}d\varphi~,
\end{eqnarray}
and
\begin{eqnarray}\label{Upsilonpolar}
\Upsilon_{polar}(v,\theta ,\varphi ) =\frac{{ - \mu_C }}{{2\sqrt 2 }}{\Lambda _ + }(H_1^{( + )}\cos \psi  + \frac{{2Q}}{{\mu_{C} r}}\Phi ){P_l}(\theta )~.
\end{eqnarray}
Now by virtue of equation \ref{Upsilonaxial}, \ref{Upsilonpolar} and \ref{current} easily the null 4-current can be obtained. For the outgoing waves one has to consider the coordinate $u=r_*-t$ and $\phi_2$ instead of $\phi_0$ and the procedure is as same as ingoing one.\\
Here based on the definition of energy momentum tensor in the N-P tetrad formalism, i.e. $${\mathop {T_{pq}}\limits^t } = {F_{pr}}F_q^r - \frac{1}{4}{\eta _{pq}}{F_{\tilde p\tilde q}}{F^{\tilde p\tilde q}}$$  we can receive the following equations for ingoing and outgoing solutions, $t$ on the top of the letters refers tetrad formalism,
\begin{eqnarray}\label{EMTENSOR-FLAT}
\begin{array}{l}
\mathop {{T_{11}}}\limits^t  =  - 2{\phi _0}\phi _0^ * ;\,\mathop {{T_{13}}}\limits^t  =  - 2{\phi _0}\phi _1^ * ;\,\mathop {{T_{12}}}\limits^t  + \mathop {{T_{30}}}\limits^t  =  - 4{\phi _1}\phi _1^ * \\
\mathop {{T_{23}}}\limits^t  =  - 2{\phi _1}\phi _2^ * ;\,\mathop {{T_{22}}}\limits^t  =  - 2{\phi _2}\phi _2^ * ;\,\mathop {{T_{33}}}\limits^t  =  - 2{\phi _0}\phi _2^ *~.
\end{array}
\end{eqnarray}
Obviously for the ingoing solutions the terms $\mathop {{T_{23}}}\limits^t $, $\mathop {{T_{22}}}\limits^t $ and $\mathop {{T_{33}}}\limits^t $ will be removable.
Also from the relation between bend space-time and tetrad basis $$\mathop {{T_{_{pq}}}}\limits^t  = e_p^ae_q^b\mathop {{T_{ab}}}\limits^B~, $$ and by virtue Eq.(\ref{ContrBasis}) we obtain
\begin{eqnarray}\label{EMTENSOR-BEND}
\begin{array}{l}
{\mathop {{T_{11}}}\limits^B} = {e^{2\psi }}\mathop {{T_{11}}}\limits^t ;\,\,~,\\
{\mathop {T_{13}}\limits^B} = ({e^{\psi  + {\mu _3}}}\mathop {{T_{13}}}\limits^t  - {q_3}{e^{2\psi }}\mathop {{T_{11}}}\limits^t )~,\\
{\mathop {T_{12}}\limits^B} = ({e^{\psi  + {\mu _2}}}\mathop {{T_{12}}}\limits^t  - {q_2}{e^{2\psi }}\mathop {{T_{11}}}\limits^t )~,\\
{q_3}{\mathop {T_{10}}\limits^{\,\,B}} + {\mathop {T_{30}}\limits^{\,\,B}} = {e^{{\mu _3} - \nu }}\mathop {{T_{30}}}\limits^t~.
\end{array}
\end{eqnarray}

\section{Appendix}

To receive the final solutions we have used some definitions and abbreviations in the context in which can be introduced as follows, for more details we refer the reader to \cite{ChandraPapers-RN paperg,ChandraPapers-RN paper,ChandraPapers-RN papera,ChandraPapers-RN paperb,ChandraPapers-RN paperc,ChandraPapers-RN paperd,ChandraPapers-RN papere,ChandraPapers-RN paperf}. Based on the turtle  coordinate and following S. Chandrasekhar works introduced in  \cite{Chandrasekhar01,Chandrasekhar02,Chandrasekhar03} we used $\,{\Lambda ^2}\,{Z^{( - )}} = {V^{( - )}}{Z^{( - )}}$, and
\[\begin{array}{l}
\frac{d}{{d{r_ * }}}\, = \frac{\Delta }{{{r^2}}}\frac{d}{{dr}},\,{\Lambda _ \pm } = \frac{d}{{d{r_ * }}} \pm i\sigma ,\,{\Lambda ^2} = {\Lambda _ + }{\Lambda _ - } = \\
{\Lambda _ - }{\Lambda _ + } = \frac{{{d^2}}}{{d{r^2}}} + i{\sigma ^2}~,\\
{r_ * } = \int {\frac{{{r^2}}}{\Delta }} dr = r + \frac{{r_ + ^2}}{{{r_ + } - {r_\_}}}\lg \left| {{r_\_} \times {r_ + }} \right| - \frac{{r_ - ^2}}{{{r_ + } - {r_\_}}}\lg \left| {{r_\_} \times {r_ - }} \right|~.
\end{array}\]
For different components of magnetic field one has
\[\begin{array}{l}
{B_{23}} = \frac{{ - Q{\mu _C}}}{{{r^2}}}H_1^{( + )} - \frac{{2{Q^2}{e^\nu }}}{{{r^3}}}\Phi \\
\,{B_{03}} = \frac{{ - Q{\mu _C}}}{{{r^2}}}H_{1,r}^{( + )} - \frac{{2{Q^2}{e^\nu }}}{{{r^4}\varpi }}(nrH_2^{( + )} + Q{\mu _C}H_1^{( + )}) + \\
\frac{{2{Q^2}{e^{ - \nu }}}}{{{r^6}}}(2{Q^2} + {r^2} + 3Mr)~,
\end{array}\]

where
\[\begin{array}{l}
H_2^{( + )} = \frac{r}{n}X - \frac{{{r^2}}}{\varpi }(L + X - {B_{23}}),\,\\
H_1^{( + )} = \frac{{ - 1}}{{Q\mu }}({r^2}{B_{23}} + 2\frac{{{Q^2}}}{r}(\frac{r}{n}X - H_2^{( + )})),\\
\Phi  = \int {(nrH_2^{( + )} + Q\mu_C H_1^{( + )})\frac{{{e^{ - \nu }}}}{{\varpi r}}} dr\, .
\end{array}\]
 One has to consider these definitions also to Eq.(\ref{Zerrili}). Now for the polar part of the perturbations and to separate the parameters $r$ and $\theta$ we considered these definitions
 \[\begin{array}{l}
\delta \nu  = N(r){P_l}(\theta ),\,\delta {\mu _2} = L(r){P_l}(\theta ),\,\\
\delta {\mu _3} = (T(r){P_l}(\theta ) + V(r){P_{l,\theta }}(\theta )\cot \theta )\\
\delta {F_{02}} = \frac{{{r^2}{e^{2\nu }}}}{{2Q}}{B_{02}}(r){P_l}(\theta ),\,{F_{03}} =  - \frac{{r{e^\nu }}}{{2Q}}{B_{03}}(r){P_{l,\theta }}(\theta ),\,\\
{F_{23}} =  - i\sigma \frac{{r{e^{ - \nu }}}}{{2Q}}{B_{23}}(r){P_{l,\theta }}(\theta )~.
\end{array}\]
And by virtue of Xanthopoulos works for axial portion of the perturbations one obtains
\[\begin{array}{l}
X = \frac{{n{e^\nu }}}{r}\Phi  + \frac{n}{r}H_2^{( + )}\, ,\\
L = \frac{{{e^\nu }}}{{{r^3}}}(3Mr - 4{Q^2})\Phi  - \frac{{(nrH_2^{( + )} + Q{\mu _C}H_1^{( + )})}}{{{r^2}}}\, ,\\
N = \frac{{{e^\nu }}}{{{r^2}}}(M - \frac{r}{\Delta }({M^2} - {Q^2} + {({r^2}\sigma )^2}))\Phi  + 2\frac{{n{e^{2\nu }}}}{\varpi }H_2^{( + )}\\
 + \frac{{(nrH_2^{( + )} + Q{\mu _C}H_1^{( + )})}}{{r{\varpi ^2}}}\{ {e^{2\nu }}[\varpi  - 2nr - 3M] - (n + 1)\varpi \} \\
 - \frac{{{e^{2\nu }}}}{\varpi }{(nrH_2^{( + )} + Q{\mu _C}H_1^{( + )})_{,r}}~.
\end{array}\]
\\
Additionally one will find the following relations useful
\[\begin{array}{l}
\frac{1}{{{{\sin }^2}\theta }}\frac{{dC_{l + 2}^{ - 3/2}(\theta )}}{{d\theta }} =  - \frac{3}{{\mu _C^2 + 2}}\frac{{d{P_l}(\theta )}}{{d\theta }}\\
{\mu _C}{X_j} =  \mp {q_j}Z_i^{( \pm )} + \mu _C^2\frac{{{r^4}}}{\Delta }(1 + \frac{{{q_j}}}{{\mu _C^2r}}){\Lambda _ + }Z_i^{( + )}~.
\end{array}\]
\section{Conclusion}
Finding an exact solution for an FFF accretion disk surrounding a central masses, for instance, a  BH was the main result of doing this work. In fact, up to now, almost solutions which have introduced in the literature were numerical and (or) non-exact solutions.  In finding these solutions we have considered a charged BH which the perturbations cause to asymmetry and consequently the rotation, as the aforementioned central mass and by virtue of FFF electrodynamics' conditions, the 4-current of such a setup has been obtained. In regards to performing our calculations we considered the well-known NP formalism and by means of the congruences of the tangent space,  we considered the 4-current and tangent basis in the same direction. Therefore, one can visualize a way to explain the ingoing, or, outgoing observed jets for a BH comparing the observational results. Also as same as the EHT results, our exact solutions can describe the  Blandford- Znajek mechanism as well.

\section*{Acknowledgments}
HS thanks A. Starobinsky for very constructive discussions about perturbations during Helmholtz International Summer School  2019 in Russia. He is grateful to G. Ellis,  and UCT for arranging his short visit, and for enlightening discussions about cosmological fluctuations and perturbations for both large and local scales. He also thanks  T. harko and H. Firouzjahi for constructive discussions about inflation and perturbations. His special thanks go to his wife E. Avirdi for her patience during our stay in South Africa.

\end{document}